\documentclass{emulateapj}

\usepackage{aas_macros} 
\usepackage[dvips]{color}
\newcommand{\ve}{\varepsilon}
\newcommand{\ff}{\mathrm{ff}}

\begin{document}

\title{
Can Gamma-Ray Burst Jets Break Out the First Stars?  
}

\author{
Yudai Suwa\altaffilmark{1,2} and Kunihito Ioka\altaffilmark{3}}

\altaffiltext{1}{Department of Physics, School of Science, the
  University of Tokyo, 7-3-1 Hongo, Bunkyo-ku,Tokyo 113-0033, Japan}
\altaffiltext{2}{Yukawa Institute for Theoretical Physics, Kyoto
  University, Oiwake-cho, Kitashirakawa, Sakyo-ku, Kyoto, 606-8502,
  Japan}
\altaffiltext{3}{KEK Theory Center and the Graduate University for
  Advanced Studies (Sokendai), 1-1 Oho, Tsukuba 305-0801, Japan }

\email{suwa@yukawa.kyoto-u.ac.jp}

\begin{abstract} 
We show that a relativistic gamma-ray burst (GRB) jet can potentially
pierce the envelope of very massive first generation star (Population
III; Pop III) by using the stellar density profile to estimate both
the jet luminosity (via accretion) and its penetrability.  The jet
breakout is possible even if the Pop III star has a supergiant
hydrogen envelope without mass loss, thanks to the long-lived powerful
accretion of the envelope itself.  While the Pop III GRB is estimated
to be energetic $E_{\gamma,\mathrm{iso}}\sim 10^{55}$ erg, the
supergiant envelope hides the initial bright phase into the cocoon
component, leading to a GRB with a long duration $\sim 1000(1+z)$ sec
and an ordinary isotropic luminosity $\sim 10^{52}$ erg s$^{-1}$
($\sim 10^{-9}$ erg cm$^{-2}$ s$^{-1}$ at redshift $z\sim 20$).  The
neutrino-annihilation is not effective for Pop III GRBs because of a
low central temperature, while the magnetic mechanism is viable.  We
also derive analytic estimates of the breakout conditions, which are
applicable to various progenitor models.  The GRB luminosity and
duration are found to be very sensitive to the core and envelope mass,
providing possible probes of the first luminous objects at the end of
the high redshift dark ages.
\end{abstract}


\section{Introduction}

The ancient era of the first generation stars (Population III; Pop
III) -- the end of the dark age -- is still an unexplored frontier in
the modern cosmology \citep{bark01,brom04,ciar05}.  The first star
formation from the metal free gas has a crucial influence on the
subsequent cosmic evolution by producing ionizing photons and heavy
elements.  Although the theoretical studies are recently developed by
the numerical simulations, the faint Pop III objects are difficult to
observe even with the future technology.

Gamma-Ray Bursts (GRBs) are potentially powerful probes of the Pop III
era.  In fact, GRB 090423 got the highest redshift $z=8.2$ ever seen
\citep[e.g.,][]{tanv09,salv09,chan10}, beyond any quasars or galaxies
and previous GRB 080913 at $z=6.7$ \citep{grei09} and GRB 050904 at
$z=6.3$ \citep{kawa06,tota06}.  The GRBs are presumed to manifest the
gravitational collapse of a massive star -- a collapsar -- to a black
hole with an accretion disk, launching a collimated outflow (jet) with
a relativistic speed \citep{macf99}.  The massive stars quickly die
within the Pop III era.  The GRBs, the most luminous objects in the
Universe, are detectable in principle out to redshifts $z\sim 100$
\citep{lamb00}, while their afterglows are observable up to $z\sim 30$
\citep{ciar00,gou04,ioka05,toma10}.  As demonstrated in GRB 050904 by
Subaru \citep{kawa06,tota06}, the GRBs can probe the interstellar
neutral fraction with the Lyman $\alpha$ red damping wing
\citep{mira98}, the metal enrichment, and the star formation rate
\citep{tota97,kist09}.  In the future, we may also investigate the
reionization history \citep{ioka03,inou04}, the molecular history
\citep{inou07}, the equation of state of the Universe
\citep{scha07,yone04} and the extragalactic background light
\citep{oh01,inou10,ferm10}.

The first stars are predicted to be predominantly very massive $\ga
100 M_{\odot}$ \citep{abel02,brom02b}.  The mass scale is roughly
given by the Jeans mass (or Bonnor- Ebert mass) when the isothermality
breaks (i.e., only through the cooling function) and hence seems
robust \citep[but see also][]{turk09,clar10}.  The central part
collapses first to a tiny ($\sim 0.01 M_{\odot}$) protostar, followed
by the rapid accretion of the surrounding matter to form a massive
first star \citep{omuk03,yosh08}.  The stars with $140$--$260
M_{\odot}$ are expected to undergo the pair-instability supernovae
without leaving any compact remnant behind, while those above $\sim
260 M_{\odot}$ would collapse to a massive ($\sim 100 M_{\odot}$)
black hole with an accretion disk, potentially leading to scaled-up
collapsar GRBs
\citep{frye01,hege03a,suwa07a,suwa07b,suwa09a,komi10,mesz10}. The Pop
III GRB rate would be rare $\sim 0.1$--$10$ yr$^{-1}$ but within reach
\citep[e.g.,][]{brom06,naoz07}.  These GRBs also marks the formation
of the first black holes, which may grow to the supermassive black
holes (BHs) via merger or accretion \citep{mada01}.

However, the zero-metal stars could have little mass loss by the line
driven wind \citep{kudr02}, and thereby have a large ($R_* \sim
10^{13}$ cm) hydrogen envelope at the end of life (red supergiant
(RSG) phase).  Especially for Pop III stars, the mass accretion
continues during the main sequence phase, so that the chemically
homogeneous evolution induced by rapid rotation
\citep[e.g.,][]{yoon05} might not work \citep{ohku09}.  Their extended
envelopes may suppress the emergence of relativistic jets out of their
surface even if such jets were produced \citep{matz03}.  The observed
burst duration $T\sim 100$ s, providing an estimate for the lifetime
of the central engine, suggests that the jet can only travel a
distance of $\sim c T \sim 10^{12}$ cm before being slowed down to a
nonrelativistic speed.  This picture is also supported by the
nondetections of GRBs associated with type II supernovae.
Nevertheless, this may not apply to the Pop III GRBs because the
massive stellar accretion could enhance the jet luminosity and
duration and therefore enable the jet to break out the first stars.

In this paper, we discuss the jet propagation in the first stars using
the state-of-the-art Pop III stellar structure calculated by
\cite{ohku09} (\S~\ref{sec:progenitor}) to estimate the jet luminosity
via accretion (\S~\ref{sec:progenitor},\ref{sec:jet}) and to predict
the observational main characters of the Pop III GRBs, such as energy
and duration.  We adopt the analytical approach that reproduces the
previous numerical simulations to see the dependences on the yet
uncertain stellar structure (\S~\ref{sec:analytic} for analytical
estimates) and to avoid simulations over many digits.  We determine
the jet head speed that is decelerated by the shock with the stellar
matter (\S~\ref{sec:penetration}).  The shocked matter is wasted as a
cocoon surrounding the jet before the jet breakout (\S~\ref{sec:GRB}),
like in the context of active galactic nuclei \citep{bege89}.  We
treat both the jet luminosity and its penetrability with the same
stellar structure consistently for the first time.

\section{Progenitor structure}\label{sec:progenitor}

In this paper we employ three representative progenitors; Pop III
star, Walf-Rayet (WR) star, and RSG. These stars correspond to
progenitors of Pop III GRBs, ordinary GRBs, and core-collapse
supernovae without GRBs, respectively.  The WR stars have no hydrogen
envelope, which is a preferred condition for a successful jet break
\citep{matz03} and consistent with the observational evidence of
GRB-SN Ibc association \citep{woos06}.

\begin{figure}[tbp]
\centering
\includegraphics[width=.45\textwidth]{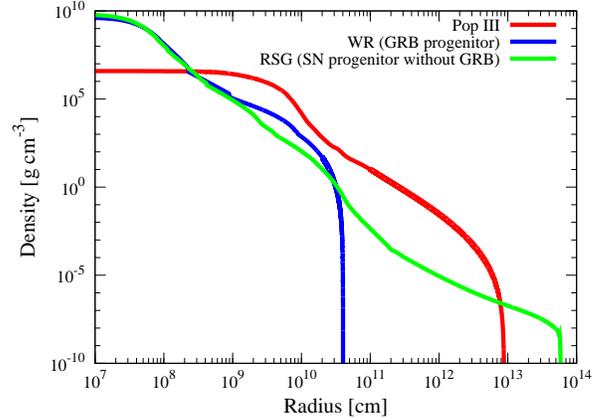}
\caption{Density profiles of investigated models. Red, blue, and green
  lines correspond to Pop III star ($M=915M_\odot$), Walf-Rayet star
  (WR; GRB progenitor, $M=16M_\odot$), and red supergiant (RSG; SN
  progenitor without GRB, $M=15M_\odot$), respectively. Pop III and
  RSG have a hydrogen envelope, which expands to a large radius, while
  WR has only a core.}
\label{fig:density}
\end{figure}

The density profiles of investigated models are shown in Fig.
\ref{fig:density}. Red line shows the density profile of Pop III star
with 915 $M_\odot$, model Y-1 of \citet{ohku09}.  Blue indicates the
GRB progenitor with 16 $M_\odot$, model 16TI of \citet{woos06b}.
Green line represents the progenitor of ordinary core-collapse
supernovae with 15 $M_\odot$, s15.0 of \citet{woos02}.  The density
profiles are roughly divided into two parts: core and envelope.  The
GRB progenitor (WR star) does not have hydrogen envelope, while Pop
III and RSG keep their envelope so that these stars experience the
envelope expansion triggered by core shrinkage after the main
sequence.

Because the exact stellar surface is difficult to calculate for the
simulation of stellar evolution (K. Nomoto, private communication), we
numerically solve the equation of hydrostatic equilibrium,
\begin{equation}
\frac{\partial P}{\partial r}=-\frac{GM_r}{r^2}\rho, 
\end{equation}
for the outermost layer of stars, where $P$ is the pressure, $r$ is
the radius from the center of the star, $G$ is the gravitational
constant, $M_r$ is the mass inside $r$, and $\rho$ is the density,
respectively. We employ the polytropic equation of state,
$P=K\rho^{\gamma}$, where $K$ is the coefficient depending on the
microphysics and $\gamma$ is the adiabatic index. Here, we use the
constant value of $K$, fitting just outside the core.  The surfaces of
stars are determined by the point with $P=0$.

We can calculate the accretion rate, $\dot M$, using these density
profiles. The accretion timescale of matter at a radius $r$ to fall to
the center of the star is roughly equal to the free-fall timescale,
\begin{equation}
t_\ff\approx \sqrt{\frac{r^3}{GM_r}}.
\end{equation} 
Then we can evaluate the accretion rate at the center as $\dot
M=dM_r/dt_\ff$.  Note that our estimation neglects the effect of
rotation \citep[e.g.,][]{kuma08,pern10}.  However the rotation law
inside the star is very uncertain. Even thought a rotationally
supported disk is formed, the accretion time is roughly $\sim
\alpha^{-1}=10(\alpha/0.1)^{-1}$ times $t_\mathrm{ff}$, where $\alpha$
is the standard dimensionless viscosity parameter \citep{kuma08}.  In
addition, the jet production mechanism is also unknown so that we
introduce an efficiency parameter to connect the (free-fall) mass
accretion rate and jet luminosity, which will be normalized by the
observed GRBs in the next section. This parameter would contain the
information of both the rotation rate and the jet production
efficiency.

\begin{figure}[tbp]
\centering
\includegraphics[width=.45\textwidth]{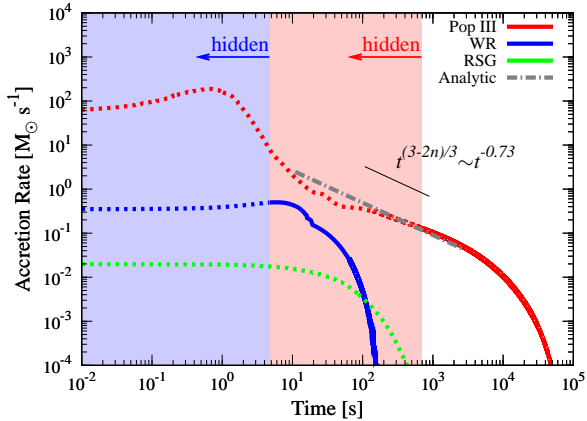}
\caption{Accretion rates as a function of time. Red, blue, and green
  lines show Pop III, WR, and RSG, respectively. Dotted regions
  represent that the jet propagates inside the star, while the solid
  regions correspond to the time after the jet breakout for the
  magnetic jet model. Solid lines give information of observables
  (e.g., duration and energetics of GRB). On the other hand, dotted
  regions show the hidden energy inside the star that goes into the
  nonrelativistic cocoon component.  The gray dot-dashed line
  represents the analytic model in Eq. (\ref{eq:L_iso}). The black
  line shows $t^{(3-2n)/3}\sim t^{-0.73}$ as a reference, where $n\sim
  2.6$ is a parameter for the density profile (an effective polytropic
  index of the envelope; see \S\ref{sec:analytic}).}
\label{fig:acc_rate}
\end{figure}

In Fig. \ref{fig:acc_rate}, the mass accretion rates of investigated
models are shown. The origin of time in this figure is set at the time
of BH mass (central mass) being $3 M_\odot$ (3.4 sec after the onset
of collapse for WR, for instance).  The accretion rate should be
related to the activity of the central engine, and that of Pop III
stars is much larger than the other progenitors.  Therefore, the GRBs
of Pop III stars are expected to be more energetic than ordinary GRBs
if Pop III stars could produce GRBs. However, it is nontrivial that
the GRB jets can breakout the Pop III stars.  In
\S\ref{sec:penetration}, we discuss the jet propagation and capability
of successful jet break.  The colored regions in this figure show the
hidden regions by the stellar interior where the jet propagates inside
the star so that the high energy photons can not be observed.

\section{Jet Models}\label{sec:jet}

In this study we employ the {\it collapsar} model, which is a widely
accepted scenario for the central engine of long GRBs. In this
scenario a black hole accompanied by the stellar collapse produces a
relativistic jet, which is strongly suggested by observations.  The
greatest uncertainty in this scenario is the mechanism for converting
the accretion energy or BH rotation energy into the directed
relativistic outflows.  There are mainly two candidates of the jet
production in the vicinity of the central engine: neutrino
annihilation and magnetohydrodynamical (MHD) mechanisms including
Blandford-Znajek process \citep{blan77}, which converts the BH
rotation energy into the Poynting flux jet via magnetic fields.
Although there are plentiful studies about these mechanisms
\citep[e.g.,][]{poph99,dima02,prog03b,mcki06}, we have no concrete
consensus for the available energy injection rate from the central
engine into the jet. Therefore, we employ two simple models for jet
producing mechanisms.  We assume that the jet injection luminosity can
be written with functions of the accretion rate, $\dot M$. The models
used in this study is basically written in $\dot M$ or $\dot M^2$. The
accretion-to-jet efficiency are given by the GRB observables with the
WR model in \S\ref{sec:GRB}.  More detailed expressions are following:

(A) $L\sim \dot M$ model (MHD mechanism): A jet is driven by magnetic
fields that are generated by accreting matter\footnote{Although the
  existence of strong magnetic field in the Pop III stars is unclear,
  there are several studies on the magnetic field amplification
  \citep[e.g.,][]{sur10}. Here we assume that the strong magnetic
  field can be generated at the vicinity of the BH.}.  In this case
the jet injection luminosity is given by $L_j=\eta \dot M c^2$, where
$\eta$ is the efficiency parameter.  In \cite{komi10},
$\eta=0.05/\alpha\beta$, where $\beta$ is the so-called plasma beta
($\beta=8\pi P/B^2$ with $B$ being magnetic field), and we do not know
the reliable values for both $\alpha$ and $\beta$ in the collapsar
scenario. Therefore we parameterize these parameters with $\eta$
simultaneously.

(B) $L\sim\dot M^2$ model (neutrino-annihilation mechanism): A jet is
driven by annihilation of neutrinos ($\nu\bar\nu\to e^-e^+$), which
are copiously radiated by ``hyperaccretion flow'' \citep{macf99}.  As
for neutrino-annihilation process, the jet injection luminosity is
written as $L_j=\zeta \dot M^{9/4} M_\mathrm{BH}^{-3/2}$
\citep{zala10}, where $\zeta$ is the efficiency parameter including
the information of accretion disk, e.g., the spectrum of emitted
neutrinos and geometry of disk.

\section{Penetration of stellar envelope}\label{sec:penetration}

In this section, we consider the propagation of the jet head in the
progenitor star.  If a relativistic jet ($\Gamma_j\gg 1$) strikes the
stellar matter, two shocks are formed: a forward shock (FS) that
accelerates the external material to a Lorentz factor $\Gamma_h$, and
a reverse shock (RS) that decelerates the head of the jet to
$\Gamma_h$.  Balancing the energy density behind the FS ($P_f$) with
that above the RS ($P_r$), one can obtain the Lorentz factor of the
jet head.  As for the ultra-relativistic case ($\Gamma_h\gg 1$),
$P_f=\frac{4}{3}\Gamma_h^2\rho c^2$ and
$P_r=\frac{4}{3}(\frac{\Gamma_j}{2\Gamma_h})^2 n_jm_pc^2$, where
$n_j=L_\mathrm{iso}/4\pi r^2\Gamma_j^2c$ is the jet proper proton
density with $L_\mathrm{iso}$ being the isotropic luminosity of a jet
and $m_p$ is the proton mass, while for nonrelativistic case
($\Gamma_h\approx 1$), $P_f=\frac{\gamma+1}{2}\rho\beta_h^2c^2$ and
$P_r=\frac{4}{3}\Gamma_j^2n_jm_p c^2$, where $\beta_h$ is the velocity
of FS in unit of the speed of light, $c$. These equations leads
following relations: ultra-relativistic one, $\Gamma_h\sim
L_\mathrm{iso}^{1/4}r^{-1/2}\rho^{-1/4}$ \citep{mesz01}, and
nonrelativistic one, $\beta_h\sim L_\mathrm{iso}^{1/2}
r^{-1}\rho^{-1/2}$ \citep{waxm03}. Here we combine these equations
empirically as follows:
\begin{eqnarray}
\beta_h\Gamma_h^2\approx 18
\left(\frac{L_\mathrm{iso}}{10^{52}\mathrm{erg~s^{-1}}}\right)^{1/2} 
\left(\frac{r}{10^{12}\mathrm{cm}}\right)^{-1}\nonumber\\
\times
\left(\frac{\rho}{10^{-7}\mathrm{g~cm^{-3}}}\right)^{-1/2}.
\label{eq:betagamma}
\end{eqnarray}
This approximation leads the same relation with \cite{waxm03} for the
nonrelativistic case ($\Gamma_h\approx 1$) and agree with
\citet{mesz01} to within 40\% for the ultra-relativistic case
($\beta_h\approx 1$).  The crossing time of the FS is also given by
\begin{equation}
t_h\approx \frac{r}{\Gamma_h^2\beta_h c}.
\label{eq:th}
\end{equation}
As for the relativistic FS, the crossing time is much shorter than the
light crossing time due to its large Lorentz factor
\cite[see][]{mesz01b}.

Combining Eqs. (\ref{eq:betagamma}) and (\ref{eq:th}), we obtain the
necessary isotropic jet luminosity for the FS to reach the radius $r$
as
\begin{eqnarray}
L_\mathrm{iso}\approx 3\times 10^{52}
\left(\frac{r}{10^{12}\mathrm{cm}}\right)^4
\left(\frac{\rho}{10^{-7}\mathrm{g~cm^{-3}}}\right)
\nonumber\\
\times
\left(\frac{t}{1\mathrm{s}}\right)^{-2}~\mathrm{erg~s^{-1}}.
\label{eq:L52}
\end{eqnarray}
If the jet luminosity decreases slower than $t^{-2}$, the jet
luminosity can achieve this value at the late phase. We can follow the
evolution of the FS by equating $L_j$ and Eq. (\ref{eq:L52}) including
the correction of the jet opening angle, $\theta_j$ (i.e.,
$L_j=L_\mathrm{iso}\theta_j^2/2$).

We note that the accreting gas from the surrounding to the progenitor
star is negligible for the jet breakout since the density of the
accreting gas is low, $\rho = \dot M/4\pi r^2 v \sim 5\times 10^{-12}$
g cm$^{-3}\ (\dot M/10^{-2}\ M_{\odot}\ {\rm yr}^{-1})
\left(r/10^{13}\ {\rm cm}\right)^{-2} \left(v/10^{8}\ {\rm cm}\ {\rm
  s}^{-1}\right)^{-1}$.

\section{GRB and cocoon}\label{sec:GRB}

In this section, we divide the energetics of the jet into two
components: GRB emitter (relativistic component) and cocoon
(nonrelativistic component). When the Lorentz factor of the FS,
$\Gamma_h$, is smaller than $\theta_j^{-1}$, the shocked material may
escape sideways and form the cocoon \citep{matz03}, which avoids the
baryon loading problem. With this scenario, the injected energy before
the shock breakout goes to the energy of cocoon and that after
breakout goes to the GRB emitter. Therefore, we can calculate the
energy budget of the GRB emitter and cocoon after the determination of
the jet breakout time, $t_b$. We define $t_b$ as the maximum time
obtained by Eq. (\ref{eq:L52}).

First of all, we determine the accretion-to-jet conversion efficiency
(depending on the mechanism) using the ordinary GRB progenitor (WR) to
make the total energy of GRB emitter $E_\mathrm{tot}=10^{52}$
erg\footnote{Note that $E_\mathrm{tot}$ is not the total energy of
  gamma rays because there must be a conversion from the jet kinetic
  energy to gamma rays. Though the efficiency of conversion is
  unclear, it is typically the order of 0.1. Therefore, we employ
  $E_\mathrm{tot}=10^{52}$ erg that could lead the true gamma-ray
  energy of GRB $\sim 10^{51}$ erg.}.  In this estimation we assume
that the half opening angle of the jet $\theta_j=5^\circ$.  A
successful GRB requires following two conditions; i) The jet head
reaches the stellar surface.  ii) The velocity of the jet head,
$\beta_h$, should be larger than that of the cocoon, $\beta_c$
\citep{matz03,toma07}.

As for $L\sim\dot M$ model, the results of the WR case are
$L_j=1.1\times 10^{51}(\dot M/M_\odot~\mathrm{s}^{-1})$ erg s$^{-1}$,
i.e.,
\begin{eqnarray}
\eta=\frac{L_j}{\dot M c^2}
\approx 6.2\times 10^{-4},
\end{eqnarray}
and $t_\mathrm{b}=4.7$ s.  For $L\sim \dot M^2$ model, $L_j=76\times
10^{51}$ $(\dot
M/M_\odot~\mathrm{s}^{-1})^{9/4}(M_\mathrm{BH}/M_\odot)^{-3/2}$ erg
s$^{-1}$ and $t_\mathrm{b}=2.8$ s\footnote{This luminosity shows a
  similar value with Eq. (22) of \cite{zala10} because if
  $M_\mathrm{BH}=3M_\odot$, $L_j\approx 16\times 10^{51}$ erg
  s$^{-1}$.}. We estimate the expected duration of the burst with the
period during which 90 percent of the burst's energy is emitted,
$T_{90}$.  Both models reproduce the typical duration of burst of
$\sim 10$ s (see Table \ref{tab:model}). The energy of cocoon
(injected energy before the shock breakout) is smaller than that of
the GRB emitter. The isotropic {\it kinetic} energy of the GRB emitter
is $\sim 10^{54}$ erg.

Next, we apply the above scheme and jet luminosity (e.g., the same
$\eta$ and $\zeta$) to the RSG (progenitor of supernovae without GRBs)
and find that RSG can not produce GRB.  This is because the jet head
is slower than the cocoon.  As for WR with $L\sim\dot M$ model,
$\beta_h\sim R_*/(ct_b) \sim 0.3$ and
$\beta_c\sim\sqrt{E_c/(Mc^2)}\sim 0.01$, where $E_c\sim 2\times
10^{51}$ erg is the energy of the cocoon (see Table 1) and $M\sim 10
M_\odot$ is the stellar mass, hence $\beta_h>\beta_c$.  On the other
hand, $\beta_h\sim 0.007$ and $\beta_c\sim 0.01$,
i.e. $\beta_h\lesssim \beta_c$, for the RSG.  Thus the morphology of
the shock wave is almost spherical and the jet cannot break out the
stellar surface with a small opening angle.  In addition, the FS
cannot reach the stellar surface with $L\sim\dot M^2$ model for the
RSG.  Therefore our scheme is consistent with observations of GRB-SN
Ibc connection.

Finally, we calculate the evolution of the jet head for the case of
Pop III star (see Table \ref{tab:model}).  We find that the $L\sim\dot
M^2$ model does {\it not} produce GRB because the FS stalls inside the
envelope due to rapidly decreasing jet luminosity (so-called ``failed
GRB''). On the other hand, the $L\sim \dot M$ model can supply enough
energy for a jet to penetrate the envelope and produce a GRB.  Since
$\beta_h\sim 0.4$ and $\beta_c\sim 0.08$ in this model, the
relativistic jet can penetrate the stellar envelope with a small
opening angle and produce a successful GRB.  The total energy of the
GRB jet (injected energy after breakout) is $\sim 45$ times larger
than the ordinary GRB and the duration is much longer ($T_{90}\sim
1000$ s).  In addition, we estimate the minimum $\eta$ for the
successful breakout, which is $\eta\approx 3.4\times 10^{-5}$.  This
is 20 times smaller than that of the normal GRB.  In this case
$\beta_h\sim 0.03$ and $\beta_c\sim 0.008$. Below this value, the jet
head cannot reach the stellar surface.

The accretion of the envelope (not core) is very important for the Pop
III GRB.  Although the envelope is mildly bounded by the gravitational
potential (because $\gamma\approx 1.38 \sim 4/3$) and easily escapes
when it is heated by the shock, the timescale of cocoon passing in the
envelope ($\sim R_*/(c\beta_c)\sim$ 3000 s) is longer than $t_b$.
Therefore, the envelope accretion can last till the jet breakout and
our conclusion about the penetrability of the relativistic jet is not
changed.

It should be noted that the opening angle of the jet could not be
constant during the propagation phase. Due to the additional
collimation by the gas pressure, $\theta_j$ becomes smaller as the jet
propagates \citep[e.g., ][]{zhan03,mizu06,mizu10}. The smaller
$\theta_j$ leads the larger $L_\mathrm{iso}(=2L_j/\theta_j^2)$ so that
our constant $\theta_j$ is a conservative assumption for the jet
breakout.

\section{Analytical dependences on parameters}\label{sec:analytic}
 
The hydrogen envelope could be reduced by the mass loss, which is one
of the most uncertain processes in the stellar evolution.  Even in the
zero-metal stars, the synthesized heavy element could be dredged up to
the surface, and might induce the line driven wind.  The stellar
luminosity could also exceed the Eddington luminosity of the envelope.
The stellar pulsation might also blow away the envelope dynamically.
So we analytically estimate the dependence on the envelope mass in the
following.

The density profile of the envelope can be written as 
\begin{equation}
\rho(r) \approx \rho_1 \left(\frac{R_*}{r}-1\right)^{n},
\label{eq:rhoenv}
\end{equation}
where $n$ is a constant \citep{matz99}.  This profile is exact if the
enclosed mass is constant (i.e., the envelope mass is negligible
compared with the core mass) and the equation of state is polytropic,
in which case $n=(\gamma-1)^{-1}$ is a polytropic index.  We have
$n=3/2$ for efficiently convective envelopes since the adiabatic index
is $\gamma=5/3$ for the ideal monoatomic gas, and $n=3$ for radiative
envelopes of constant opacity $\kappa$ since $P\propto \rho^{4/3}$ is
derived from the relations, $P_{\gamma}=(L/L_{\rm Edd}) P$, $P \propto
\rho T$, $P_{\gamma} \propto T^4$, with a constant luminosity-mass
ratio $L/M$ where $L_{\rm Edd}=4\pi G M c/\kappa > L$ is the Eddington
luminosity.  We can fit well the Pop III envelope in
Fig.~\ref{fig:density} with $n \approx 2.6$.

Using the profile in equation (\ref{eq:rhoenv}), we can estimate the
envelope mass as
\begin{equation}
M_\mathrm{env} = \int_{R_c}^{R_*} \rho(r) 4\pi r^2 dr \propto
\frac{\rho_1 R_*^3}{3-n} 
\sim \frac{\rho_c R_c^{n} R_*^{3-n}}{3-n}, 
\end{equation}
where $\rho_c\equiv\rho(R_c)\approx\rho_1(R_*/R_c)^{n}$ is the
envelope density just above the core and $R_c$ is the core radius.
The core density is higher than $\rho_c$ to proceed with the nuclear
burning.  Since the density enhancement is determined by the
difference of the ignition temperature, which is not sensitive to
other parameters, we assume that the core mass is given by $M_c\propto
\rho_c R_c^3$, so that
\begin{equation}
M_\mathrm{env}\propto \frac{M_c R_c^{n-3}R_*^{3-n}}{3-n}.  
\end{equation}
Then the stellar radius can be written as a function of the core
radius, the core mass, and the envelope mass as follows:
\begin{eqnarray}
R_* \sim 10^{13}\ {\rm cm}
\left(\frac{R_c}{10^{10}~\mathrm{cm}}\right)
\left(\frac{M_c}{400M_\odot}\right)^{-2.5}
\nonumber\\
\times
\left(\frac{M_\mathrm{env}}{500\ M_{\odot}}\right)^{2.5},
\label{eq:R_*}
\end{eqnarray}
where we use $n=2.6$ (see Appendix for $n$ dependences).  Therefore,
the stellar radius has a strong dependence on the envelope mass.  If
the envelope mass is smaller than $\sim 50 M_\odot$, the stellar
radius is almost the core radius, $R_c\sim 10^{10}$ cm.

Next, we consider the jet breakout time. Since the accretion time of
the core ($r\lesssim 10^{10}$ cm) is $\sim 4$ s, the envelope
accretion is important for the successful breakout if $t_b$ is longer
than this timescale.  Using $t_\ff\sim\sqrt{r^3/GM_r}$, we can
evaluate the envelope accretion rate as
\begin{equation}
{\dot M} = \frac{dM_r/dr}{dt_\ff/dr} 
\propto \rho_1 M_c^{(3-n)/3} R_*^{n} t^{(3-2n)/3}
\label{eq:dotM}
\end{equation}
with the approximation
of $M_r-M_c = \int_{R_c}^{r} \rho(r') 4\pi r'^2 dr'
\ll M_c \approx 400M_\odot$, which is valid for $r\lesssim 10^{12}$ cm
(corresponding to $t_\ff\lesssim 3000$ s).  
Combining with Eqs. (\ref{eq:L52}), (\ref{eq:R_*}), (\ref{eq:dotM}) and 
$L_j=\eta \dot M c^2 = L_{\rm iso} \theta_j^2/2$, 
the breakout time is
given by
\begin{eqnarray}
t_b &\sim& 700~\mathrm{s}
\left(\frac{\eta}{10^{-3}}\right)^{-0.79}
\left(\frac{\theta_j}{5^\circ}\right)^{1.6}
\left(\frac{R_c}{10^{10}~\mathrm{cm}}\right)^{1.1}
\nonumber\\
&&\times
\left(\frac{M_c}{400 M_\odot}\right)^{-2.9}
\left(\frac{M_\mathrm{env}}{500 M_\odot}\right)^{2.8},
\label{eq:t_b}
\end{eqnarray}
where we use $n=2.6$ (see Appendix for $n$ dependences) and put
$\rho=\rho_1$ and $r=R_*$ in Eq.~(\ref{eq:L52}).  If $\eta$ is very
small ($\lesssim 10^{-5}$) and $t_b$ is longer than the free-fall
timescale of the outermost part of star ($\sim 10^5$ s), the jet
cannot penetrate the star.  When $\eta\approx 10^{-5}$, $\beta_h\sim
R_*/t_b\sim 0.003$ and $\beta_c\sim\sqrt{\eta}\sim 0.003$ so that the
shock wave also propagates almost spherically.  Similarly, the RSG
case in Fig.~\ref{fig:acc_rate} also has too long breakout time since
$M_c \sim 4 M_{\odot}$ and $M_{\rm env} \sim 11 M_{\odot}$.

The jet luminosity after the breakout ($t> t_b$) is given by
\begin{eqnarray}
L_\mathrm{iso}(t) &\sim& 5\times 10^{52}
\left(\frac{\eta}{10^{-3}}\right) 
\left(\frac{\theta_j}{5^\circ}\right)^{-2}
\left(\frac{R_c}{10^{10}~\mathrm{cm}}\right)^{-0.4} \nonumber\\ 
&&\times
\left(\frac{M_c}{400M_\odot}\right)^{1.1}
\left(\frac{t}{700~\mathrm{s}}\right)^{-0.73} \mathrm{erg~s}^{-1},
\label{eq:L_iso}
\end{eqnarray}
where we use $n=2.6$ (see Appendix for $n$ dependences) and $M_c
\propto \rho_c R_c^3 \sim \rho_1 (R_*/R_c)^n R_c^3$.  Interestingly
the dependence of $L_{\rm iso}(t)$ on the envelope mass is only
through the time, $t$.  The relativistic jet emitted after the
breakout will produce GRB prompt emission and the jet luminosity
decreases with time as $t^{-(2n-3)/3}\sim t^{-0.73}$ in this case.
Fig. \ref{fig:acc_rate} shows this dependence with the gray dot-dahsed
line, which well reproduces the numerical result (red line) for $10$ s
$\lesssim t \lesssim 3000$ s.  The parameter dependences just after
the breakout (most luminous time; $t=t_b$) are given by
Eqs. (\ref{eq:t_b}) and (\ref{eq:L_iso}) as
\begin{eqnarray}
L_\mathrm{iso}(t=t_b)&\sim&
 5\times 10^{52}
\left(\frac{\eta}{10^{-3}}\right) ^{1.6}
\left(\frac{\theta_j}{5^\circ}\right)^{-3.2} 
\left(\frac{R_c}{10^{10}~\mathrm{cm}}\right)^{-1.2}
\nonumber\\&&\times
\left(\frac{M_c}{400M_\odot}\right)^{3.2} 
\left(\frac{M_\mathrm{env}}{500M_\odot}\right)^{-2.0}
~\mathrm{erg~s}^{-1},
\label{eq:Lisotb}
\end{eqnarray}
where we use $n=2.6$ (see also Appendix).

The active duration of the central engine ($\approx t_b+T_{90}$) is
determined by the accretion timescale at $r\sim 10^{12}$ cm, where the
density gradient gets larger because
$\rho(r)\propto(R_*/r-1)^{n}$. The region of $r\lesssim 10^{12}$ cm
has $\rho(r)\propto (R_*/r)^{n}$ so that Eqs. (\ref{eq:t_b}) and
(\ref{eq:L_iso}) are valid, while they are not appropriate for
$r\gtrsim 10^{12}$ cm ($M_r\gtrsim M_c+0.4M_\mathrm{env}$) due to the
existence of the stellar surface, which leads the fast decrease of the
accretion rate (compare the red and gray lines in
Fig. \ref{fig:acc_rate}).  So, the duration can be calculated by
$t_\mathrm{ff}$ at $r\sim 0.1R_*$ with Eq. (\ref{eq:R_*}) as
\begin{eqnarray}
&&t_\mathrm{ff}(r=0.1R_*)
\sim
3000
\left(\frac{R_c}{10^{10}~\mathrm{cm}}\right)^{1.5}
\left(\frac{M_\mathrm{c}}{400M_\odot}\right)^{-3.8}
\nonumber\\
&&\times
\left(\frac{M_\mathrm{env}}{500M_\odot}\right)^{3.8}
\left(\frac{M_c+0.4 M_\mathrm{env}}{600M_\odot}\right)^{-0.5}
~\mathrm{s},
\label{eq:tff}
\end{eqnarray}
where we use $n=2.6$ (see also Appendix). This is consistent with the
result $t_b+T_{90}\sim 2200$ s in the previous section.  At this time,
the isotropic luminosity in Eq.~(\ref{eq:L_iso}) is given by
\begin{eqnarray}
&&L_\mathrm{iso}[t=t_\mathrm{ff}(r=0.1R_*)]
\sim 2\times 10^{52}
\left(\frac{\eta}{10^{-3}}\right)
\left(\frac{\theta_j}{5^\circ}\right)^{-2}
\nonumber\\
&&\times
\left(\frac{R_c}{10^{10}~\mathrm{cm}}\right)^{-1.5}
\left(\frac{M_c}{400M_\odot}\right)^{3.9}
\left(\frac{M_\mathrm{env}}{500M_\odot}\right)^{-2.8}
\nonumber\\
&&\times
\left(\frac{M_c+0.4M_\mathrm{env}}{600M_\odot}\right)^{0.37}
\mathrm{erg~s^{-1}},
\label{eq:Ltff}
\end{eqnarray}
where we use $n=2.6$ (see also Appendix).  We note that the
observables in Eqs.~(\ref{eq:Lisotb}), (\ref{eq:tff}) and
(\ref{eq:Ltff}) carry information of the stellar structure.

\section{Summary and Discussion}
 
We have investigated the jet propagation in the very massive
Population III stars assuming the accretion-to-jet conversion
efficiency of the observed normal GRBs.  We find that the jet can
potentially break out the stellar surface even if the Pop III star has
a massive hydrogen envelope thanks to the long-lasting accretion of
the envelope itself.  Even if the accretion-to-jet conversion is less
efficient than the ordinary GRBs by a factor of $\sim 20$, the jet
head can penetrate the stellar envelope and produce GRBs.  Although
the total energy injected by the jet is as large as $\sim10^{54}$ erg,
more than half is hidden in the stellar interior and the energy
injected before the breakout goes into the cocoon component.  The
large envelope accretion can activate the central engine so that the
duration of Pop III GRB is very long if the hydrogen envelope exists.
As a result, the luminosity of Pop III GRB is modest, being comparable
to that of ordinary GRBs.

Considering the Pop III GRB at redshift $z$, the duration in
Eq.~(\ref{eq:tff}) is
\begin{equation}
T_\mathrm{GRB}=T_{90}(1+z)\approx 30000~\mathrm{s}~
\left(\frac{1+z}{20}\right), 
\end{equation}
which is much longer than the canonical duration of GRBs, $\sim 20$ s.
The total isotropic-equivalent energy of Pop III GRB is
\begin{equation}
E_{\gamma,\mathrm{iso}}=\ve_\gamma E_\mathrm{iso}\approx 1.2
\times10^{55} \left(\frac{\ve_\gamma}{0.1}\right)\ {\rm erg}, 
\end{equation}
where $\ve_\gamma$ is the conversion efficiency from the jet kinetic
energy to gamma rays (see Table \ref{tab:model}).  It should be noted
that this value is comparable to the largest $E_{\gamma,\mathrm{iso}}$
ever observed, $\approx 9\times 10^{54}$ erg for GRB 080916C
\citep{abdo09b}. This value is smaller than the estimation of
\cite{komi10,mesz10} because we consider the hidden (cocoon)
component.  Since the large isotropic energy is stretched over the
long duration, the expected flux just after the breakout is not so
bright,
\begin{equation}
F=\frac{\ve_\gamma L_\mathrm{iso}}{4\pi r_L^2}\sim 10^{-9}\ {\rm
  erg}\ {\rm cm}^{-2} \ {\rm s}^{-1},
\end{equation}
where $r_L$ is the luminosity distance, which is smaller than the {\it
  Swift} Burst Array Telescope (BAT) sensitivity, $\sim 10^{-8}$ erg
cm$^{-2}$ s$^{-1}$. However, there must be a large variety of the
luminosity as ordinary GRBs so that more luminous but rare events
might be observable by BAT.  Although the cocoon component has a large
energy, the velocity is so low that it is also difficult to observe.
If the cocoon component interacts with the dense wind or ambient
medium, it might be observable.  This is an interesting future work.

The above discussions strongly depend on the envelope mass because the
stellar radius is highly sensitive to the envelope mass.  We derive
analytical dependences on the model parameters in
\S~\ref{sec:analytic}, which are applicable to a wide variety of
progenitor models.  According to the analytical estimates, the smaller
envelope leads to the shorter duration and the larger observable
luminosity.  If $M_\mathrm{env}\lesssim 50 M_\odot$, $R_*\sim R_c\sim
10^{10}$ cm (see Eq.(\ref{eq:R_*})), $t_b\sim 2$ s (obtained using
numerical model as in \S\ref{sec:GRB}), and $T_{90}\sim
t_\mathrm{ff}\sim
4(M/400M_\odot)^{-1/2}(R_c/10^{10}\mathrm{cm})^{3/2}$ s; Both $t_b$
and $T_{90}$ might be extended by a factor of $\alpha^{-1}$ due the
the rotation.  This is much shorter than the GRBs from Pop III stars
with massive envelopes.  The mass accreting after the breakout is
$\sim 200 M_\odot$ so that $E_{\gamma,\mathrm{iso}}\sim 3\times
10^{54}(\ve_\gamma/0.1)$ erg can be emitted by gamma rays after
breakout. The luminosity just after the breakout is
$L_\mathrm{iso}\sim 10^{54}$ erg s$^{-1}$, i.e., $F\sim 10^{-7}\ {\rm
  erg}\ {\rm cm}^{-2}\ {\rm s}^{-1}$, which is much brighter than the
case with the massive envelope and observable with the Swift BAT,
while the duration is very short ($\sim 2$ s at the source frame).

The matter entrainment from the envelope is also crucial for the
fireball dynamics and the GRB spectra \citep[e.g.,][]{ioka10}.  Since the
envelope of the PopIII star is different from that of present-day
stars, the GRB appearance is also likely distinct from the observed
ones.  This is also an interesting future work.

Since we do not have any conclusive central engine scenario, we employ
two popular mechanisms in this paper, that is, the magnetic model ($L
\sim \dot M$) and neutrino-annihilation model ($L \sim {\dot M}^2$).
The difference between these models is the dependence of the mass
accretion rate. We find that the neutrino-annihilation model cannot
penetrate the Pop III stellar envelope, so that no GRB occurs, which
is consistent with \cite{frye01,komi10,mesz10}.

\begin{table*}
\caption{Model Summary}
\begin{tabular}{l|cclccccc}
\hline
Model & Final Mass & Radius & Mechanism & $t_\mathrm{b}$ & Energy of GRB emitter & Energy of Cocoon & $T_{90}$ & $E_{\mathrm{iso}}$\\
(Reference) & [$M_\odot$] & [$10^{11}$ cm] & & [s] & [$10^{52}$ erg] & [$10^{52}$ erg] & [s] & [$10^{54}$ erg] \\
\hline
WR & 14 & 0.4 & MHD & 4.7 & 1.0 & 0.23 & 49 & 2.6\\
\citep{woos06b}  & & & Neutrino & 2.8 & 1.0 & 0.42 & 10 & 2.6\\
\hline
Pop III & 915 & 90 & MHD & 690 & 45 & 57 & 1500 & 120\\
\citep{ohku09} && & Neutrino & --- & --- & --- & --- & ---\\
\hline
RSG & 13 & 600 & MHD & ($5\times 10^5$) &  (0.88) & (0.16) & ($8.4\times 10^6$)\\
\citep{woos02}  & & & Neutrino & --- &  --- & --- & --- & ---\\
\hline
\end{tabular}
\tablecomments{The final masses of WR and RSG are not the same mass as
  the initial mass of progenitors ($15M_\odot$ and $16M_\odot$)
  because of mass loss due to stellar wind, while Pop III stars do not
  have mass loss so that the final mass is $915 M_\odot$.  $t_b$ is
  the time of shock breakout from our calculations (see
  \S~\ref{sec:GRB}).  The energy of GRB emitter is the {\it kinetic}
  energy of relativistic jet after the breakout. The energy of the
  cocoon is the injected energy before the breakout. $E_\mathrm{iso}$
  is the isotropic energy of GRB emitter, corrected by the half
  opening angle of the jet, $\theta_j=5^\circ$. The parenthetic values
  for RSG are just for references because the cocoon velocity is
  larger than GRB emitter so that the GRB is not generated.}
\label{tab:model}
\end{table*}

\acknowledgments 

We would like to thank A. Heger and T. Ohkubo for providing the
progenitor models and P. M\'esz\'aros and K. Omukai for valuable
discussions.  This study was supported in part by the Japan Society
for Promotion of Science (JSPS) Research Fellowships (YS) and by the
Grant-in-Aid from the Ministry of Education, Culture, Sports, Science
and Technology (MEXT) of Japan (No.19047004, 21684014, 22244019).

\appendix

\section{Dependence on the density index $n$}
Here we explicitly show the dependences on $n$ in Eqs.~(\ref{eq:R_*}),
(\ref{eq:t_b}), (\ref{eq:L_iso}), (\ref{eq:Lisotb}), (\ref{eq:tff})
and (\ref{eq:Ltff}) as follows,
\begin{eqnarray}
R_* &\propto& R_c M_c^{-\frac{1}{3-n}} [(3-n) M_{\rm env}]^{\frac{1}{3-n}},
\\
t_b &\propto& 
\eta^{-\frac{3}{9-2n}}
\theta_j^{\frac{6}{9-2n}}
R_c^{\frac{3(4-n)}{9-2n}}
M_c^{-\frac{n^2-9n+21}{(3-n)(9-2n)}}
\left[(3-n)M_{\rm env}\right]^{\frac{3(4-n)}{(3-n)(9-2n)}},
\\
L_{\rm iso}(t) &\propto&
\eta \theta_j^{-2} R_c^{-(3-n)} M_c^{\frac{6-n}{3}} t^{-\frac{2n-3}{3}},
\\
L_{\rm iso}(t=t_b) &\propto&
\eta^{\frac{6}{9-2n}}
\theta_j^{-\frac{12}{9-2n}}
R_c^{-\frac{15-4n}{9-2n}}
M_c^{\frac{2n^2-16n+33}{(3-n)(9-2n)}}
\left[(3-n)M_\mathrm{env}\right]^{-\frac{(4-n)(2n-3)}{(3-n)(9-2n)}},
\\
t_{\rm ff}(r=0.1 R_*) &\propto&
R_c^{\frac{3}{2}}
M_c^{-\frac{3}{2(3-n)}}
[(3-n)M_\mathrm{env}]^{\frac{3}{2(3-n)}}
(M_c+0.4M_\mathrm{env})^{-\frac{1}{2}},
\\
L_\mathrm{iso}[t=t_\mathrm{ff}(r=0.1R_*)] 
&\propto&
\eta\theta_j^{-2}
R_c^{-\frac{3}{2}}
M_c^{\frac{2n^2-12n+27}{6(3-n)}}
[(3-n)M_\mathrm{env}]^{\frac{3-2n}{2(3-n)}}
(M_c+0.4M_\mathrm{env})^{\frac{2n-3}{6}}.
\end{eqnarray}
We note that the total energy is proportional to
\begin{eqnarray}
t_{\rm ff}(r=0.1 R_*) L_\mathrm{iso}[t=t_\mathrm{ff}(r=0.1R_*)]
\propto
\eta\theta_j^{-2}
M_c^{\frac{3-n}{3}}
[(3-n)M_\mathrm{env}]
(M_c+0.4M_\mathrm{env})^{-\frac{3-n}{3}}.
\end{eqnarray}
if $n<3$ (i.e., $L_{\rm iso}(t)$ is shallower than $t^{-1}$).

\end{document}